\documentstyle[12pt,epsf]{article}
\thicklines
\begin{document}

\vspace*{3cm}
\begin{center}
{\large\bf  Prospects for the $\sf B_c$ Studies at LHCb}\\

\vspace*{1cm}
{\bf I.P.Gouz, V.V.Kiselev, A.K.Likhoded, V.I.Romanovsky,}\\
 and {\bf O.P.Yushchenko}\\

\vspace*{0.7cm}
IHEP, Protvino, Russia\\

~~~
\end{center}
\begin{abstract}
We discuss the motivations and perspectives  for the studies of the mesons of 
the $(bc)$ family at LHCb.
The description of production and decays at LHC energies is given in details.
The event yields, detection efficiencies, and background conditions for several
$B_c$ decay modes at LHCb are estimated.
\end{abstract}

\newpage

\section{Introduction}

The $B_c$ meson is the ground state of $\bar{b}c$ system which in many respects
is an intermediate between charmonium and bottonium systems. However because the
$B_c$ mesons carry flavour, they provide a window for studying the heavy-quark
dynamics that is very different from those provided by $c\bar{c}$- and
$b\bar{b}$-quarkonia.

The first observation of approximately 20 $B_c$ events in the $B_c\to J\Psi
l\nu$ decay mode by the CDF collaboration \cite{cdf} demonstrates the
possibility of the experimental study of the $B_c$ meson.

The $\bar{b}c$ states have rich spectroscopy of the  orbital and
angular-momentum exitations. Below the threshold of the decay into $B-D$ pair,
one can expect 16 extremely narrow states which cascadely decay into the ground
pseudoscalar state with mass of about 6.3 GeV by radiating 
photons and pion pairs.
The annihilation decays can occure due to weak interactions only and, hence, are
suppressed for excited levels.

The production mechanism for $\bar{b}c$ system differs in an essential way from
that for $\bar{b}b$ system, because two heavy quark-antiquark pairs must be
created in a collision. While the $\bar{b}b$ pair can be created in the parton
processes $q\bar{q},\; gg\to b\bar{b}$ at the order of $\alpha_S^2$, the lowest
order mechanism for the creating of $\bar{b}c$ system is at least of
$\alpha_S^4$: $q\bar{q},\; gg\to (\bar{b}c) b\bar{c}$, and gluon-gluon
contribution dominates at Tevatron and LHC energies. At LHC with the luminosity
of about {$\cal L$} $ = 10^{34} \mbox{cm}^{-2}\mbox{s}^{-1}$ one could expect 
around
$5\times 10^{10}$ $B_c$ events per year.
At Tevatron energy the expected yield should be at least one order smaller.

The weak decays of $B_c$ mesons are attractive due to presence of both channels:
a) the $b$-quark decay with the $c$ quark as a spectator, and b) the
$c$-constituent decay with $b$ as a spectator.
 In addition the weak annihilation
contribution into decay channels is quite visible (around 10\%).

The dominant contribution into $B_c$ life-time ($\tau_{B_c} \sim 0.5$ ps) comes
from the $c$-quark decays ($\sim$ 70\%) while the $b$-quark decays and weak
annihilation add about 20\%\ and 10\%\ respectively.

The accurate measurement of the $B_c$ life-time can provide us with the information
on both the masses of charm and beauty quarks and the normalization point of
non-leptonic weak Lagrangian in the $B_c$ decays.

The experimental study of the semileptonic decays and the extraction of the
decay form-factors can test the spin symmetry derived in NRQCD and HQET
approaches. The measurement of the branching fractions for semileptonic and
hadronic modes can provide an information about
 the parameters of weak Lagrangian and hadronic
matrix elements determined by the non-perturbative effects due to quark
confinement.

\section{ The mass spectrum of the $\bf (\bar b c)$ family}

%\noindent
The most accurate estimates of $(\bar b c)$ masses \cite{ger,eq} can be
obtained in the framework of nonrelativistic potential models based on the
NRQCD expansion over both $1/m_Q$ and $v_{rel}\to 0$ \cite{bra}.

The uncertainty of evaluation is about 30 MeV. The reason is
the following. The potential models \cite{pot} were justified for the well
measured masses of charmonium and bottomonium. So, the potentials with
various global behaviour, i.e. with the different $r\to \infty$ and
$r\to 0$ asymptotics, have the same form in the range of mean distances
between the quarks in the heavy quarkonia at $0.2 < r < 1$ fm \cite{eic}.
The observed  regularity in the distances between the excitation levels
are approximately flavor-independent. The latter is exact for the
logarithmic potential (the Feynman--Hell-Mann theorem), where the
average kinetic energy of quarks $T$ is a constant value independent of
the excitation numbers (the virial theorem) \cite{log}. 
A slow dependence of the
level distances on the reduced mass can be taken into account by the use of
the Martin potential (power law:
$V(r)=A(r/r_0)^a+C$, $a\ll 1$) \cite{mart}, wherein the predictions are
in agreement with the QCD-motivated
Buchm\" uller-Tye potential  with the account for the two-loop 
evolution of the coupling constant at short distances \cite{bt}.

So, one gets the picture of $(\bar b c)$ levels which is very close to the
texture of charmonium and bottomonium.
The difference is the {\sf jj}-binding instead of the
{\sf LS} one.

The spin-dependent perturbation of the potential includes the
contribution of the effective one-gluon exchange (the vector part)
as well as the scalar confining term \cite{fein}.
\begin{eqnarray}
   V_{SD}(\vec{r}) & = &\biggl(\frac{\vec{L}\cdot\vec{S}_c}{2m_c^2} +
\frac{\vec{L}\cdot\vec{S}_b}{2m_b^2}\biggr)\;
\biggl(-\frac{dV(r)}{rdr}+\frac{8}{3}\;\alpha_s\;\frac{1}{r^3}\biggr) 
+ \nonumber \\
~ & ~ & +\frac{4}{3}\;\alpha_s\;\frac{1}{m_c
m_b}\;\frac{\vec{L}\cdot\vec{S}}{r^3}
+\frac{4}{3}\;\alpha_s\;\frac{2}{3m_c m_b}\;
\vec{S}_c\cdot\vec{S}_b\;4\pi\;\delta(\vec{r}) \label{3} \\
~ & ~ & +\frac{4}{3}\;\alpha_s\;\frac{1}{m_c m_b}\;(3(\vec{S}_c\cdot\vec{n})\;
(\vec{S}_b\cdot\vec{n}) - \vec{S}_c\cdot\vec{S}_b)\;\frac{1}{r^3}\;,
\;\;\vec{n}=\frac{\vec{r}}{r}\;.\nonumber
\end{eqnarray}

The model-dependent value of effective $\alpha_s$ \cite{eq} can
be extracted from the data on the splitting in the charmonium
$$
M(J/\Psi)-M(\eta_c) = \alpha_s \frac{8}{9m_c^2} |R(0)|^2\approx 117 \;
{\rm MeV.}
$$
We take into account the renormalization-group dependence
of $\alpha_s$ at the one-loop accuracy by means of introduction of
the quarkonium scale \cite{ger}
$$
\mu^2 = \langle {\bf p}^2\rangle =2 \langle T\rangle m_{red}.
$$
The estimated difference between the masses of basic pseudoscalar state
and its vector excitation \cite{ger} is equal to 
$$
M(B_c^{*+})-M(B_c^+)=65\pm 15\; {\rm MeV.}
$$
The mass of the ground state \cite{ger} equals 
\begin{equation}
M(B_c^+)=6.25\pm 0.03\; {\rm GeV,}
\end{equation}
which is in agreement with the CDF measurements $M(B_c) = 6.4\pm 0.19$ GeV
\cite{cdf}.

\begin{figure}[hbt]
\epsfxsize=10cm \epsfbox{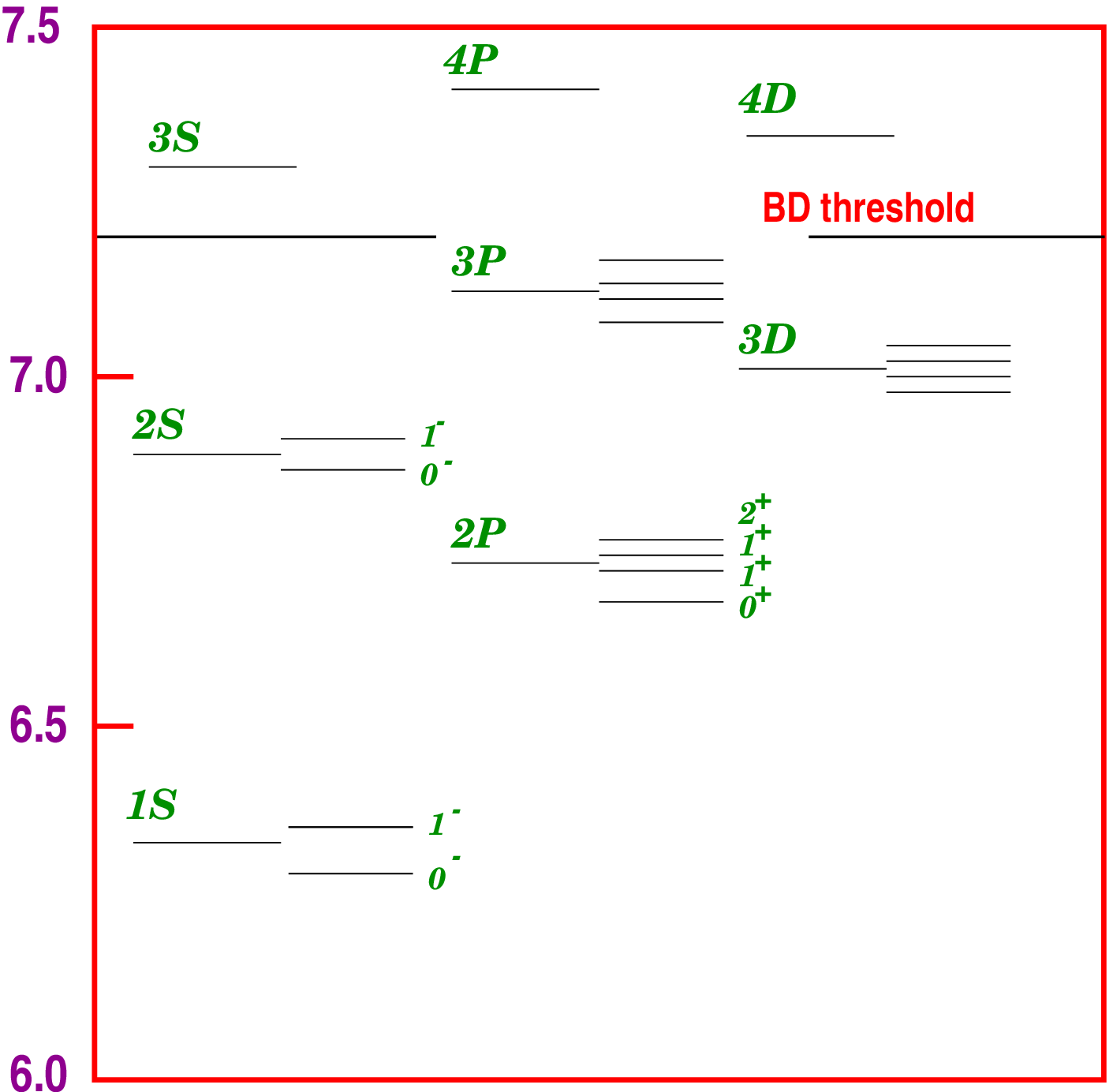}

\vspace*{-9.5cm} \hfill
\begin{tabular}{||l|c|c||}
\hline
state & Martin  & BT \\
\hline
$1^1S_0$    & 6.253       & 6.264    \\
$1^1S_1$    & 6.317       & 6.337    \\
$2^1S_0$    & 6.867       & 6.856    \\
$2^1S_1$    & 6.902       & 6.899    \\
$2^1P_0$    & 6.683       & 6.700    \\
$2P\; 1^+$  & 6.717       & 6.730    \\
$2P\; 1'^+$ & 6.729       & 6.736    \\
$2^3P_2$    & 6.743       & 6.747    \\
$3^1P_0$    & 7.088       & 7.108    \\
$3P\; 1^+$  & 7.113       & 7.135    \\
$3P\; 1'^+$ & 7.124       & 7.142    \\
$3^3P_2$    & 7.134       & 7.153    \\
$3D\; 2^-$  & 7.001       & 7.009    \\
$3^5D_3$    & 7.007       & 7.005    \\
$3^3D_1$    & 7.008       & 7.012    \\
$3D\; 2'^-$ & 7.016       & 7.012    \\
\hline
\end{tabular}

\vspace*{1.cm} 

\caption{The mass spectrum of $(\bar b c)$ with account for the
spin-dependent splittings.}
%\label{f1k}
\end{figure}

\subsection{ Radiative transitions}

 The bright feature of the $(\bar b c)$ family is that there are no
annihilation
decay modes due to the strong interaction. So, the excitations, in a cascade
way,
decay into the ground state with the emission of {\sf photons} and {\sf
pion-pion} pairs.

The formulae for the {\sf E1}-transitions are slightly modified.
\begin{eqnarray}
\Gamma(\bar nP_J\to n^1S_1 +\gamma) & = & \frac{4}{9}\;\alpha_{{\rm
em}}\;Q^2_{{\rm eff}}\;
       \omega^3\;I^2(\bar nP;nS)\;w_J(\bar nP) \;,\nonumber\\
\Gamma(\bar nP_J\to n^1S_0 +\gamma) & = & \frac{4}{9}\;\alpha_{{\rm
em}}\;Q^2_{{\rm eff}}\;
       \omega^3\;I^2(\bar nP;nS)\;(1-w_J(\bar n P)) \;,\nonumber\\
\Gamma(n^1S_1\to \bar n P_J +\gamma) & = & \frac{4}{27}\;\alpha_{{\rm
em}}\;Q^2_{{\rm eff}}\;
       \omega^3\;I^2(nS;\bar n P)\;(2J+1)\;w_J(\bar n P)\;, \label{7} \\
\Gamma(n^1S_0\to \bar n P_J +\gamma) & = & \frac{4}{9}\;\alpha_{{\rm
em}}\;Q^2_{{\rm eff}}\;
       \omega^3\;I^2(nS;\bar n P)\;(2J+1)\;(1-w_J(\bar n P))\;,\nonumber\\
\Gamma(\bar n P_J\to n D_{J'} +\gamma) & = & \frac{4}{27}\;\alpha_{{\rm
em}}\;Q^2_{{\rm eff}}\;
       \omega^3\;I^2(nD;\bar n P)\;(2J'+1)\;
\nonumber\\ && w_J(\bar n P)) w_{J'}(nD)
        S_{JJ'}\;,\nonumber\\
\Gamma(n D_J\to \bar n P_{J'} +\gamma) & = & \frac{4}{27}\;\alpha_{{\rm
em}}\;Q^2_{{\rm eff}}\;
       \omega^3\;I^2(nD;\bar n P)\;(2J'+1)\;
\nonumber\\ && w_{J'}(\bar n P)) w_{J}(nD)
        S_{J'J}\;,\nonumber
\end{eqnarray}
where $\omega$ is the photon energy, $\alpha_{\rm em}$ is the electromagnetic
fine structure constant. 
In eq.(\ref{7}) one uses
\begin{equation}
   Q_{{\rm eff}}=\frac{m_c Q_{\bar b} - m_b Q_c }{m_c +m_b}\;, 
\end{equation}
where $Q_{c,b}$ are the electric charges of the quarks. For the
$B_c$ meson with the parameters from the Martin potential, one gets 
$Q_{\rm eff}=0.41$. 
$w_J(nL)$ is the probability that the spin $S=1$ 
in the $nL$ state. $S_{JJ'}$ are the statistical factors.
The $I(\bar nL;nL')$ value is expressed through the radial wave functions,
\begin{equation}
   I(\bar n L;nL') = |\int R_{\bar n L}(r) R_{nL'}(r) r^3 {\rm d}r|\;. 
\label{7aa}
\end{equation}

For the dipole magnetic {\sf M1}-transitions one has 
\begin{equation}
\Gamma(\bar n^1S_i\to n^1S_f +\gamma) =  \frac{16}{3}\;\mu^2_{{\rm
eff}}\;\omega^3\;
(2f+1)\;A_{if}^2\;,
\label{10}
\end{equation}
where 
$$ A_{if} = \int R_{\bar n S}(r) R_{nS}(r) j_0(\omega r/2) r^2\; {\rm d}r\;,
$$
and
\begin{equation}
\mu_{{\rm eff}}=\frac{1}{2}\;\frac{\sqrt{\alpha_{{\rm em}}}}{2m_c m_b}\;
(Q_c m_b - Q_{\bar b} m_c)\;.
\label{11}
\end{equation}
Note, that in 
contrast to the $\psi$  and $\Upsilon$ particles, the total width of
the $B_c^*$ meson is equal to the width of its radiative decay into
the $B_c(0^-)$ state.

Thus, below the threshold of decay into the BD-pair
the theory predicts the existence of 16
narrow $(\bar b c)$ states, which do not annihilate due to the strong
interactions, but they have the cascade radiative transitions into the ground
long-lived pseudoscalar state, the $B_c^+$ meson.

\begin{table}[hbt]
\begin{center}
\begin{tabular}{||l|r|p{29.3mm}|r||}
\hline
state & $\Gamma_{\rm tot}$, KeV  &dominant decay mode & BR, \% \\
\hline
$1^1S_1$    & 0.06       & $1^1S_0+\gamma$   & 100  \\
$2^1S_0$    & 67.8       & $1^1S_0+\pi\pi$   &  74  \\
$2^1S_1$    & 86.3       & $1^1S_1+\pi\pi$   &  58  \\
$2^1P_0$    & 65.3       & $1^1S_1+\gamma$   & 100  \\
$2P\; 1^+$  & 89.4       & $1^1S_1+\gamma$   &  87  \\
$2P\; 1'^+$ & 139.2      & $1^1S_0+\gamma$   &  94  \\
$2^3P_2$    & 102.9      & $1^1S_1+\gamma$   & 100  \\
$3^1P_0$    & 44.8       & $2^1S_1+\gamma$   &  57  \\
$3P\; 1^+$  & 65.3       & $2^1S_1+\gamma$   &  49  \\
$3P\; 1'^+$ & 92.8       & $2^1S_0+\gamma$   &  63  \\
$3^3P_2$    & 71.6       & $2^1S_1+\gamma$   &  69  \\
$3D\; 2^-$  & 95.0       & $2P\; 1^++\gamma$ &  47  \\
$3^5D_3$    & 107.9      & $2^3P_2+\gamma$   &  71  \\
$3^3D_1$    & 155.4      & $2^1P_0+\gamma$   &  51  \\
$3D\; 2'^-$ & 122.0      & $2P\; 1'^++\gamma$ & 38 \\
\hline
\end{tabular}
\end{center}
%\vspace*{2cm}
\caption{The total widths of $(\bar b c)$-states with Martin potential}
\label{t2}
\end{table}
%\newpage

\section{\boldmath $B_c$ lifetime and inclusive decay
rates \label{2}}

The $B_c$-meson decay processes can be subdivided into three classes:

1) the $\bar b$-quark decay with the spectator $c$-quark,

2) the $c$-quark decay 
with the spectator $\bar b$-quark and

3) the annihilation channel
$B_c^+\rightarrow l^+\nu_l (c\bar s, u\bar s)$, where $l=e,\; \mu,\; \tau$.

In the $\bar b \to \bar c c\bar s$ decays one separates also the
Pauli interference with the $c$-quark from the initial state. In
accordance with the given classification, the total width is the sum over the
partial widths
$$%%\begin{equation}
\Gamma (B_c\rightarrow X)=\Gamma (b\rightarrow X)
+\Gamma (c\rightarrow X)+\Gamma \mbox{(ann.)}+\Gamma\mbox{(PI)}.
$$%%\end{equation}
For the annihilation channel the  $\Gamma\mbox{(ann.)}$ width can be reliably
estimated in the framework of inclusive approach, where one takes the sum of
the leptonic and quark decay modes with account for the hard gluon corrections
to the effective four-quark interaction of weak currents. These corrections
result in the factor of $a_1=1.22\pm 0.04$. The width is expressed through the
leptonic constant of $f_{B_c}\approx 400$ MeV. This estimate of the
quark-contribution does not depend on a hadronization model, since a large
energy release of the order of the meson mass takes place. From the following
expression, one can see that the contribution by light leptons and quarks can
be neglected,
$$%%\begin{equation}
\Gamma \mbox{(ann.)} =\sum_{i=\tau,c}\frac{G^2_F}{8\pi}
|V_{bc}|^2f^2_{B_c}M m^2_i (1-m^2_i/m^2_{Bc})^2\cdot C_i\;,
\label{d3}
$$%%\end{equation}
where $C_\tau = 1$ for the $\tau^+\nu_\tau$-channel and 
$C_c =3|V_{cs}|^2a_1^2 $ for the $c\bar s$-channel.

As for the non-annihilation decays, in the approach of the  Operator
Product Expansion for the quark currents of weak decays \cite{OPEBc}, one
takes into account the $\alpha_s$-corrections to the free quark decays and uses
the quark-hadron duality for the final states. Then one considers the matrix
element for the transition operator over the bound meson state. The latter
allows one also to take into account the effects caused by the motion and
virtuality of decaying quark inside the meson because of the interaction with
the spectator. In this way the $\bar b\to \bar c c\bar s$ decay mode turns out
to be suppressed almost completely due to the Pauli interference with the charm
quark from the initial state. Besides, the $c$-quark decays with the spectator
$\bar b$-quark are essentially suppressed in comparison with the free quark
decays because of a large bound energy in the initial state. 

\begin{table}[th]
\begin{center}
\begin{tabular}{|l|c|c|c|}
\hline
$B_c$ decay mode & OPE, \%  & PM, \% & SR, \%\\
\hline
$\bar b\to \bar c l^+\nu_l$ & $3.9\pm 1.0$  & $3.7\pm 0.9$  & $2.9\pm 0.3$\\
$\bar b\to \bar c u\bar d$  & $16.2\pm 4.1$ & $16.7\pm 4.2$ & $13.1\pm 1.3$\\
$\sum \bar b\to \bar c$     & $25.0\pm 6.2$ & $25.0\pm 6.2$ & $19.6\pm 1.9$\\
$c\to s l^+\nu_l$           & $8.5\pm 2.1$  & $10.1\pm 2.5$ & $9.0\pm 0.9$\\
$c\to s u\bar d$            & $47.3\pm 11.8$& $45.4\pm 11.4$& $54.0\pm 5.4$\\
$\sum c\to s$               & $64.3\pm 16.1$& $65.6\pm 16.4$& $72.0\pm 7.2$\\
$B_c^+\to \tau^+\nu_\tau$   & $2.9\pm 0.7$  & $2.0\pm 0.5$  & $1.8\pm 0.2$\\
$B_c^+\to c\bar s$          & $7.2\pm 1.8$  & $7.2\pm 1.8$  & $6.6\pm 0.7$\\
\hline
\end{tabular}
\end{center}
\caption{The branching ratios of the $B_c$ decay modes calculated in
the framework of inclusive OPE approach, by summing up the exclusive modes in
the potential model \cite{PMK,PML} and according to the semi-inclusive
estimates in the sum rules of QCD and NRQCD \cite{KLO,KKL}. }
\label{t5}
\label{inc}
\end{table}

In the framework of exclusive approach, it is necessary to sum up widths
of different decay modes calculated in the potential models. While considering
the semileptonic decays due to the $\bar b \to \bar c l^+\nu_l$ and $c\to s
l^+\nu_l$ transitions, one finds that the hadronic final states are practically
saturated by the lightest bound $1S$-state in the $(\bar c c)$-system, i.e. by
the $\eta_c$ and $J/\psi$ particles, and the $1S$-states in the $(\bar b
s)$-system, i.e. $B_s$ and $B_s^*$, which can only enter the accessible
energetic gap. 

Further, the $\bar b\to \bar c u\bar d$ channel, for example, can be calculated
through the given decay width of $\bar b \to \bar c l^+\nu_l$ with account
for the color factor and hard gluon corrections to the four-quark 
interaction. It can be also obtained as a sum over the widths of decays 
with the $(u\bar d)$-system bound states. 

The results of calculation for the total $B_c$ width in the inclusive OPE and
exclusive PM approaches give the values consistent with each other, if one 
takes into account the most significant uncertainty related to the choice
of quark masses (especially for the charm quark), so that finally, we have 
\begin{equation}
\left.\tau[B_c^+]\right._{\mbox{\small\sc ope,\,pm}}= 0.55\pm 0.15\; \mbox{ps,}
\end{equation}
which agrees with the measured value of $B_c$ lifetime.

\begin{figure}[th]
\setlength{\unitlength}{0.5mm}
\begin{center}
\begin{picture}(130,95)
\put(0,5){\epsfxsize=120\unitlength \epsfbox{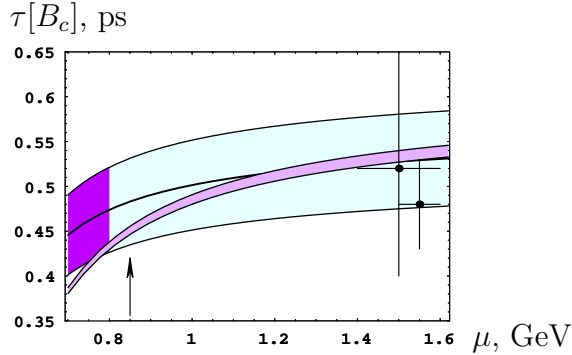}}
\put(0,92){$\tau[{B_c}] $, ps}
\put(123,7){$\mu$, GeV}
\end{picture}
\end{center}
\caption{The $B_c$ lifetime calculated in QCD sum rules versus the scale of
hadronic weak lagrangian in the decays of charmed quark. The wide shaded region
shows the uncertainty of semi-inclusive estimates, the dark shaded region is
the preferable choice as given by the lifetimes of charmed mesons. The dots
represent the values in OPE approach taken from ref. \cite{OPEBc}. The narrow
shaded region represents the result obtained by summing up the exclusive
channels with the variation of hadronic scale in the decays of beauty
anti-quark in the range of $1 <\mu_b < 5$ GeV. The arrow points to the
preferable prescription of $\mu =0.85$ GeV as discussed in \cite{KKL}.}
\label{life}
\end{figure}

The OPE estimates of inclusive decay rates agree with recent semi-inclusive
calculations in the sum rules of QCD and NRQCD \cite{KLO,KKL}, where one
assumed the saturation of hadronic final states by the ground levels in the
$c\bar c$ and $\bar b s$ systems as well as the factorization allowing one to
relate the semileptonic and hadronic decay modes. The coulomb-like corrections
in the heavy quarkonia states play an essential role in the $B_c$ decays and
allow one to remove the disagreement between the estimates in sum rules and
OPE. In contrast to OPE, where the basic uncertainty is given by the variation
of heavy quark masses, these parameters are fixed by the two-point sum rules
for bottomonia and charmonia, so that the accuracy of SR calculations for the
total width of $B_c$ is determined by the choice of scale $\mu$ for the
hadronic weak lagrangian in decays of charmed quark. We show this dependence in
Fig. \ref{life}, where $\frac{m_c}{2} < \mu < m_c$ and the dark shaded region
corresponds to the scales preferred by data on the charmed meson lifetimes.

Supposing the preferable choice of scale in the $c\to s$ decays of $B_c$ to be
equal to $\mu^2_{B_c} \approx (0.85\; {\rm GeV})^2$, putting $a_1(\mu_{B_c})
=1.20$ and neglecting the contributions caused by nonzero $a_2$ in the charmed
quark decays \cite{KKL}, in the framework of semi-inclusive sum-rule
calculations we predict
\begin{equation}
\left.\tau[B_c]\right._{\mbox{\small\sc sr}} = 0.48\pm 0.05\;{\rm ps},
\end{equation}
which agrees with the direct summation of exclusive channels calculated in the
next sections. In Fig. \ref{life} we show the exclusive estimate of lifetime,
too.

\section{Semileptonic and leptonic modes\label{5}}
\subsection{Semileptonic decays}
The semileptonic decay rates are underestimated in the QCD SR approach of ref.
\cite{QCDSRBc}, because large coulomb-like corrections were not taken into
account. The recent analysis of SR in \cite{KT,KLO,KKL} decreased the
uncertainty, so that the estimates agree with the calculations in the potential
models.%%% (see Table \ref{semi}).

The absolute values of semileptonic widths are presented in Table \ref{tp81} in
comparison with the estimates obtained in potential models. 

In practice, the most constructive information is given by the $J/\Psi$ mode,
since this charmonium is clearly detected in experiments due to the pure
leptonic decays \cite{cdf}. In addition to the investigation of various form
factors and their dependence on the transfer squared, we would like to stress
that the measurement of decay to the excited state of charmonium, i.e.
$\psi^\prime$, could answer the question on the reliability of QCD predictions
for the decays to the excited states. We see that to the moment the finite
energy sum rules predict the width of $B_c^+\to \psi^\prime l^+ \nu$ decays
in a reasonable agreement with the potential models.

\begin{table}[h]
\begin{center}
\begin{tabular}{|l|r|r|r|r|r|r|}
\hline
%Channel &
~~~~~Mode & $\Gamma$ \cite{newK} & $\Gamma$ \cite{vary} & $\Gamma$
\cite{chch} & $\Gamma$ \cite{ivanov} & $\Gamma$ \cite{ISGW2} & $\Gamma$
\cite{narod}\\
\hline
 $B_c^+ \rightarrow \eta_c e^+ \nu$                 
& 11 &11.1 &14.2  & 14 &10.4 &8.6\\
 $B_c^+ \rightarrow \eta_c \tau^+ \nu$                 
& 3.3 &  &  & 3.8 & &2.9\\
 $B_c^+ \rightarrow \eta_c^\prime e^+ \nu$                 
& 0.60 &  & 0.73  &  &0.74 & \\
 $B_c^+ \rightarrow \eta_c^\prime \tau^+ \nu$                 
& 0.050 &  &  & & & \\
 $B_c^+ \rightarrow J/\psi e^+ \nu $ 
& 28 &30.2 &34.4 & 33 & 16.5  &18\\
 $B_c^+ \rightarrow J/\psi \tau^+ \nu $ 
& 7.0 & &  & 8.4 & &5.0 \\
%{$\bar{b}$ decay} & 
 $B_c^+ \rightarrow \psi^\prime e^+ \nu $ 
& 1.94 & & 1.45 & &3.1 & \\
 $B_c^+ \rightarrow \psi^\prime \tau^+ \nu $ 
& 0.17 & & &  & & \\
 $B_c^+ \rightarrow  D^0 e^+ \nu $                  
& 0.059 & 0.049 & 0.094 & 0.26 & 0.026 & \\
 $B_c^+ \rightarrow  D^0 \tau^+ \nu $                  
& 0.032 & &  &0.14 & & \\
 $B_c^+ \rightarrow  D^{*0} e^+ \nu  $              
& 0.27 & 0.192 & 0.269 & 0.49 & 0.053 & \\
 $B_c^+ \rightarrow  D^{*0} \tau^+ \nu  $              
& 0.12 & & & 0.27 & & \\
\hline\hline
 $B_c^+ \rightarrow  B^0_s e^+ \nu  $               
& 59 &14.3 &26.6 &29 &13.8 &15 \\
%$c$ decay &
 $B_c^+ \rightarrow B_s^{*0} e^+ \nu  $    
& 65 &50.4 &44.0 &37 &16.9 &34\\
  $B_c^+ \rightarrow B^0 e^+ \nu  $                  
& 4.9 &1.14 &2.30 &2.1 & & \\
 $B_c^+ \rightarrow B^{*0} e^+ \nu  $               
& 8.5 &3.53 &3.32 &2.3 & & \\
\hline
\end{tabular}
\caption{
Exclusive widths of semileptonic $B_c^+$ decays, $\Gamma$ in
$10^{-15}$ GeV.
}
\label{tp81}
\end{center}
\end{table}

\subsection{Leptonic decays}

The dominant leptonic decay of $B_c$ is given by the $\tau \nu_\tau$ mode (see
Table \ref{inc}). However, it has a low experimental efficiency of detection
because of hadronic background in the $\tau$ decays or a missing energy.
Recently, in refs. \cite{radlep} the enhancement of muon and electron channels
in the radiative modes was studied. The additional photon allows one to remove
the helicity suppression for the leptonic decay of pseudoscalar particle, which
leads, say, to the double increase of muonic mode.

\section{Non-leptonic modes\label{6}}

In comparison with the inclusive non-leptonic widths, which can be
estimated in the framework of quark-hadron duality (see Table \ref{inc}), the
calculations of exclusive modes usually involves the approximation of
factorization \cite{fact}, which, as expected, can be quite accurate for the
$B_c$, since the quark-gluon sea is suppressed in the heavy quarkonium. Thus,
the important parameters are the factors $a_1$ and $a_2$ in the non-leptonic
weak lagrangian, which depend on the normalization point suitable for the $B_c$
decays.

The QCD SR estimates for the non-leptonic decays of charmed quark in $B_c$ give
the agreement of
results with the values predicted by the potential models is rather good for
the direct transitions with no permutation of colour lines, i.e. the class I
processes with the factor of $a_1$ in the non-leptonic amplitude determined by
the effective lagrangian. In contrast, the sum rule predictions are
significantly enhanced in comparison with the values calculated in the
potential models for the transitions with the colour permutation, i.e. for the
class II processes with the factor of $a_2$. 

Further, for the transitions, wherein the Pauli interference is significantly
involved, the class III processes, we find that the absolute values of
different terms given by the squares of $a_1$ and $a_2$ calculated in the sum
rules are in agreement with the estimates of potential models. However, we
stress that 
 we have found that due to the Pauli interference determining
the negative sign of two amplitudes with $a_1$ and $a_2$ the overall sign in
some modes should be different from those obtained in the potential models.
 Taking into account the negative value of $a_2$ with
respect to $a_1$, we see that half of decays should be
enhanced in comparison with the case of Pauli interference switched off, while
the other half is suppressed. The characteristic values of effects caused by
the Pauli interference is presented in Table \ref{Pauli}, where we put the
widths in the form
$$
\Gamma = \Gamma_0 + \Delta \Gamma,\qquad \Gamma_0=x_1\, a_1^2+x_2\, a_2^2,\quad
\Delta\Gamma =z a_1\, a_2.
$$
Then, we conclude that the Pauli interference can be straightforwardly tested
in the listed decays, wherein its significance reaches about 50\%.

\begin{table}[ht]
\begin{center}
\begin{tabular}{|l|c|}
\hline
~~~~~Mode & $\Delta\Gamma/\Gamma_0$, \% \\
\hline
$B_c^+ \rightarrow \eta_c D_s^+$             
& ~59 \\
$B_c^+ \rightarrow \eta_c D_s^{*+}$          
& -41 \\
$B_c^+ \rightarrow J/\psi D_s^+$             
& -55 \\
$B_c^+ \rightarrow J/\psi D_s^{*+}$          
& ~53 \\
$B_c^+ \rightarrow \eta_c D^+$            
& ~43\\
$B_c^+ \rightarrow \eta_c D^{*+}$            
& -47 \\
$B_c^+ \rightarrow J/\psi D^+$               
& -46 \\
$B_c^+ \rightarrow J/\psi D^{*+}$            
& ~48 \\
\hline
\end{tabular}
\caption{The effect of Pauli interference in the exclusive non-leptonic decay
widths of the $B_c$ meson with the $c$-quark as spectator at $a_1=1.18$,
$a_2=-0.22$.}
\label{Pauli}
\end{center}
\end{table}

At large recoils as in $B_c^+\to J/\Psi \pi^+(\rho^+)$, the spectator picture of
transition can be broken by the hard gluon exchanges \cite{Gers}. The spin
effects in such decays were studied in \cite{Pakh}. However, we emphasize that
the significant rates of $B_c$ decays to the P- and D-wave charmonium states
point out that the corrections in the second order of the heavy-quark velocity
in the heavy quarkonia under study could be quite essential and suppress the
corresponding decay rates, since the relative momentum of heavy quarks inside
the quarkonium if different from zero should enhance the virtuality of gluon
exchange, which suppresses the decay amplitudes.

The widths of non-leptonic $c$-quark decays
in the framework of the  sum rule are greater than those of potential models. 
In this respect we check
that our calculations are consistent with the inclusive ones. So, we sum up
the calculated exclusive widths and estimate the total width of $B_c$ meson as
shown in Fig. \ref{life}, which points to a good agreement of our calculations
with those of OPE and semi-inclusive estimates.

Another interesting point is the possibility to extract the factorization
parameters $a_1$ and $a_2$ in the $c$-quark decays by measuring the branching
ratios
\begin{eqnarray}
\frac{\Gamma[B_c^+\to B^+\bar K^0]}{\Gamma[B_c^+\to B^0 K^+]} &=& 
\frac{\Gamma[B_c^+\to B^+\bar K^{*0}]}{\Gamma[B_c^+\to B^0 K^{*+}]} = \nonumber
\\[3mm]
\frac{\Gamma[B_c^+\to B^{*+}\bar K^0]}{\Gamma[B_c^+\to B^{*0} K^+]} &=& 
\frac{\Gamma[B_c^+\to B^{*+}\bar K^{*0}]}{\Gamma[B_c^+\to B^{*0} K^{*+}]} =
\label{factk}\\[3mm]
&=& \left|\frac{V_{cs}}{V_{cd}^2}\right|^2\,\left(\frac{a_2}{a_1}\right)^2.
\nonumber
\end{eqnarray}
This procedure can give the test for the factorization approach itself.

The suppressed decays caused by the flavor changing neutral currents were
studied in \cite{rare}.

The CP-violation in the $B_c$ decays can be investigated in the same
manner as made in the charged $B$ decays. The expected  CP-asymmetry of ${\cal
A}(B_c^\pm \to J/\psi D^\pm)$ is about $4\cdot 10^{-3}$, when the corresponding
branching ratio is suppressed as $10^{-4}$ \cite{CPv}. 
The reference-triangle ideology can by applied for the model-independent
extraction of CKM-matrix angle $\gamma$. However, the corresponding branchings
are suppressed, e.g. $\mbox{Br}(B_c^+\to D_s^+ D^0) \sim 10^{-5}$.
Thus, the direct study
of CP-violation in the $B_c$ decays is practically difficult because of
low relative yield of $B_c$ with respect to ordinary $B$ mesons:
$\sigma(B_c)/\sigma(B) \sim 10^{-3}$.

Another possibility is the lepton tagging of $B_s$ in the $B_c^\pm\to B_s^{(*)}
l^\pm \nu$ decays for the study of mixing and {\sf CP}-violation in the $B_s$
sector \cite{Quigg}.

\begin{table*}[th]
\caption{Branching ratios \cite{newK} 
of exclusive $B_c^+$ decays at the fixed choice of
factors: $a_1^c =1.20$ and $a_2^c=-0.317$ in the non-leptonic decays of $c$
quark, and $a_1^b =1.14$ and $a_2^b=-0.20$ in the non-leptonic decays of $\bar
b$ quark. The lifetime of $B_c$ is appropriately normalized by $\tau[B_c]
\approx 0.45$ ps.}
\label{common}
\begin{center}
\begin{tabular}{|l|r|}
\hline
~~~~~Mode & BR, \%\\
\hline
 $B_c^+ \rightarrow \eta_c e^+ \nu$                 
 & 0.75\\
 $B_c^+ \rightarrow \eta_c \tau^+ \nu$                 
 & 0.23\\
 $B_c^+ \rightarrow \eta_c^\prime e^+ \nu$                 
 & 0.041\\
 $B_c^+ \rightarrow \eta_c^\prime \tau^+ \nu$                 
 & 0.0034\\
 $B_c^+ \rightarrow J/\psi e^+ \nu $ 
 & 1.9\\
 $B_c^+ \rightarrow J/\psi \tau^+ \nu $ 
 & 0.48\\
 $B_c^+ \rightarrow \psi^\prime e^+ \nu $ 
 & 0.132 \\
 $B_c^+ \rightarrow \psi^\prime \tau^+ \nu $ 
 & 0.011\\
 $B_c^+ \rightarrow  D^0 e^+ \nu $                  
 & 0.004 \\
 $B_c^+ \rightarrow  D^0 \tau^+ \nu $                  
 & 0.002 \\
 $B_c^+ \rightarrow  D^{*0} e^+ \nu  $              
 & 0.018  \\
 $B_c^+ \rightarrow  D^{*0} \tau^+ \nu  $              
 & 0.008 \\
%%\hline\hline
 $B_c^+ \rightarrow  B^0_s e^+ \nu  $               
 & 4.03  \\
 $B_c^+ \rightarrow B_s^{*0} e^+ \nu  $    
 & 5.06 \\
  $B_c^+ \rightarrow B^0 e^+ \nu  $                  
 & 0.34\\
 $B_c^+ \rightarrow B^{*0} e^+ \nu  $               
 & 0.58 \\
%%\hline \hline
 $B_c^+ \rightarrow \eta_c \pi^+$      
 & 0.20\\
 $B_c^+ \rightarrow \eta_c \rho^+$     
 & 0.42\\
 $B_c^+ \rightarrow J/\psi \pi^+$      
 & 0.13\\
 $B_c^+ \rightarrow J/\psi \rho^+$     
 & 0.40\\
 $B_c^+ \rightarrow \eta_c K^+ $     
 & 0.013\\
 $B_c^+ \rightarrow \eta_c K^{*+}$     
 & 0.020\\
\hline
\end{tabular}
\begin{tabular}{|l|r|}
\hline
~~~~~Mode & BR, \%\\
\hline
 $B_c^+ \rightarrow J/\psi K^+$        
 & 0.011\\
 $B_c \rightarrow J/\psi K^{*+}$       
 & 0.022\\
 $B_c^+ \rightarrow D^+ 
\overline D^{\hspace{1pt}\raisebox{-1pt}{$\scriptscriptstyle 0$}}$           
 & 0.0053\\
 $B_c^+ \rightarrow D^+ 
\overline D^{\hspace{1pt}\raisebox{-1pt}{$\scriptscriptstyle *0$}}$
 & 0.0075\\
 $B_c^+ \rightarrow  D^{\scriptscriptstyle *+} 
\overline D^{\hspace{1pt}\raisebox{-1pt}{$\scriptscriptstyle 0$}}$       
 & 0.0049\\
 $B_c^+ \rightarrow  D^{\scriptscriptstyle *+} 
\overline D^{\hspace{1pt}\raisebox{-1pt}{$\scriptscriptstyle *0$}}$
 & 0.033\\
 $B_c^+ \rightarrow D_s^+ \overline
D^{\hspace{1pt}\raisebox{-1pt}{$\scriptscriptstyle 0$}}$
 & 0.00048\\
 $B_c^+ \rightarrow D_s^+ 
\overline D^{\hspace{1pt}\raisebox{-1pt}{$\scriptscriptstyle *0$}}$
 & 0.00071\\
 $B_c^+ \rightarrow  D_s^{\scriptscriptstyle *+} \overline
D^{\hspace{1pt}\raisebox{-1pt}{$\scriptscriptstyle 0$}}$
 & 0.00045\\
 $B_c^+ \rightarrow  D_s^{\scriptscriptstyle *+} 
\overline D^{\hspace{1pt}\raisebox{-1pt}{$\scriptscriptstyle *0$}}$
 & 0.0026\\
%%\hline \hline
 $B_c^+ \rightarrow \eta_c D_s^+$             
 & 0.86\\
 $B_c^+ \rightarrow \eta_c D_s^{*+}$          
 & 0.26\\
 $B_c^+ \rightarrow J/\psi D_s^+$             
 & 0.17\\
 $B_c^+ \rightarrow J/\psi D_s^{*+}$          
 & 1.97\\
 $B_c^+ \rightarrow \eta_c D^+$            
 & 0.032\\
 $B_c^+ \rightarrow \eta_c D^{*+}$            
 & 0.010\\
 $B_c^+ \rightarrow J/\psi D^+$               
 & 0.009\\
 $B_c^+ \rightarrow J/\psi D^{*+}$            
 & 0.074\\
%%\hline \hline
 $B_c^+ \rightarrow B_s^0 \pi^+$              
 & 16.4\\
 $B_c^+ \rightarrow B_s^0 \rho^+$             
 & 7.2\\
 $B_c^+ \rightarrow B_s^{*0} \pi^+$           
 & 6.5\\
 $B_c^+ \rightarrow B_s^{*0} \rho^+$          
 & 20.2\\
\hline
\end{tabular}
\begin{tabular}{|l|r|}
\hline
~~~~~Mode & BR, \%\\
\hline
 $B_c^+ \rightarrow B_s^0 K^+$              
 & 1.06\\
 $B_c^+ \rightarrow B_s^{*0} K^+$             
 & 0.37\\
 $B_c^+ \rightarrow B_s^0 K^{*+}$             
 & --\\
 $B_c^+ \rightarrow B_s^{*0} K^{*+}$          
 & --\\
 $B_c^+ \rightarrow B^0 \pi^+$                
 & 1.06\\
 $B_c^+ \rightarrow B^0 \rho^+$               
 & 0.96\\
 $B_c^+ \rightarrow B^{*0} \pi^+$             
 & 0.95\\
 $B_c^+ \rightarrow B^{*0} \rho^+$            
 & 2.57\\
 $B_c^+ \rightarrow B^0 K^+$                  
 & 0.07\\
 $B_c^+ \rightarrow B^0 K^{*+}$               
 & 0.015\\
 $B_c^+ \rightarrow B^{*0} K^+$               
 & 0.055\\
 $B_c^+ \rightarrow B^{*0} K^{*+}$            
 & 0.058\\
%%\hline\hline
 $B_c^+ \rightarrow B^+ \overline{K^0}$       
 & 1.98\\
 $B_c^+ \rightarrow B^+ \overline{K^{*0}}$    
 & 0.43\\
 $B_c^+ \rightarrow B^{*+} \overline{K^0}$    
 & 1.60\\
 $B_c^+ \rightarrow B^{*+} \overline{K^{*0}}$ 
 & 1.67\\
 $B_c^+ \rightarrow B^+ \pi^0$             
 & 0.037\\
 $B_c^+ \rightarrow B^+ \rho^0$               
 & 0.034\\
 $B_c^+ \rightarrow B^{*+} \pi^0$             
 & 0.033\\
 $B_c^+ \rightarrow B^{*+} \rho^0$            
 & 0.09\\
 $B_c^+ \rightarrow \tau^+ \nu_\tau$            
 & 1.6\\
 $B_c^+ \rightarrow c \bar s$            
 & 4.9\\
\hline
\end{tabular}
\end{center}
\end{table*}

We present here the current status of the $B_c$ meson decays.
We have found that the various
approaches: OPE, Potential models and QCD sum rules, result in the close
estimates, while the SR as explored for the various heavy quark systems, lead
to a smaller uncertainty due to quite an accurate knowledge of the heavy quark
masses. So, summarizing, we expect that the dominant contribution to the $B_c$
lifetime is given by the charmed quark decays ($\sim 70\%$), while the
$b$-quark decays and the weak annihilation add about 20\% and 10\%,
respectively. The coulomb-like $\alpha_s/v$-corrections play an essential role
in the determination of exclusive form factors in the QCD SR. The form factors
obey the relations dictated by the spin symmetry of NRQCD and HQET with quite
a good accuracy expected.

The predictions of QCD sum rules for the exclusive decays of $B_c$ are
summarized in Table \ref{common} at the fixed values of factors $a_{1,2}$ and
lifetime. In addition to the decay channels with the heavy charmonium $J/\psi$
well detectable through its leptonic mode, one could expect a significant
information on the dynamics of $B_c$ decays from the channels with  single
heavy mesons, if an experimental efficiency allows one to extract a signal from
the cascade decays. An interesting opportunity is presented by the relations
for the ratios in (\ref{factk}), which can shed light to characteristics of the
non-leptonic decays in the explicit form.

We have found that the $\bar b$ decay to the doubly charmed states gives
$$
\mbox{Br}[B_c^+\to \bar c c\,c\bar s] \approx 3.26\%,
$$
so that in the absolute value of width it can be compared with the estimate of
spectator decay \cite{OPEBc},
\begin{eqnarray}
\left.\Gamma[B_c^+\to \bar c c\,c\bar s]\right|_{\mbox{\sc sr}} &\approx &
48\cdot
10^{-15}\,\mbox{GeV},\nonumber \\[2mm]
\left.\Gamma[B_c^+\to \bar c c\,c\bar s]\right|_{\rm spect.} &\approx & 90\cdot
10^{-15}\,\mbox{GeV}, \nonumber
\end{eqnarray}
and we find the suppression factor of about $1/2$. This result is in agreement
with the estimate in OPE \cite{OPEBc}, where a strong dependence of negative
term
caused by the Pauli interference on the normalization scale of non-leptonic
weak lagrangian was emphasized, so that at large scales one gets approximately
the same suppression factor of $1/2$, too.

To the moment we certainly state that the accurate direct measurement of $B_c$
lifetime can provide us with the information on both the masses of charmed and
beauty quarks and the normalization point of non-leptonic weak lagrangian in
the $B_c$ decays (the $a_1$ and $a_2$ factors). The experimental study of
semileptonic decays and the extraction of ratios for the form factors can test
the spin symmetry derived in the NRQCD and HQET approaches and decrease the
theoretical uncertainties in the corresponding theoretical evaluation of quark
parameters, as well as the hadronic matrix elements, determined by the
nonperturbative effects caused by the quark confinement. The measurement of
branching fractions for the semileptonic and non-leptonic modes and their
ratios can give information
 on the values of factorization parameters, which depend again
on the normalization of non-leptonic weak lagrangian. The charmed quark
counting in the $B_c$ decays is related to the overall contribution of $b$
quark decays, as well as with the suppression of $\bar b\to c\bar c \bar s$
transition because of the destructive Pauli interference, which value depends
on
the nonperturbative parameters (roughly estimated, the leptonic constant) and
non-leptonic weak lagrangian.

Thus, the progress in measuring the $B_c$ lifetime and decays could improve
the theoretical understanding of what really happens in the heavy quark decays
at all.

We point also to the papers, wherein some aspects of $B_c$ decays and
spectroscopy were studied:
non-leptonic decays in \cite{verma}, polarization effects in the radiative
leptonic decays \cite{giri}, relativistic effects \cite{Nobes}, spectroscopy in
the systematic approach of potential nonrelativistic QCD in \cite{Brambilla},
nonperturbative effects in the semileptonic decays \cite{Mannel}, exclusive and
inclusive decays of $B_c$ states into the lepton pair and hadrons \cite{Ma},
rare decays in \cite{Ivanov2}, the spectroscopy and radiative decays in
\cite{Ebert}.

\section{$\bf B_c$ production}

The $(\bar b c)$ system is a heavy quarkonium, i.e. it contains two heavy
quarks.
This determines the general features for the $B_c$ meson production in various
interactions:

{\sf 1.} Perturbative calculations for the hard associative production of
two heavy pairs of $\bar c c$ and $\bar b b$ and

{\sf 2.} A soft nonperturbative binding of nonrelativistic quarks in the
color-singlet
state can be described in the framework of potential models.

The two above conditions result in the suppression of the 
 $B_c$ yield of the order of 
  $10^{-3}$ with respect to beauty hadrons.

As was mentioned above, the consideration of mechanisms for the hadronic 
production of different spin $B_c$-states is based on the factorization of 
hard parton production of heavy quarks $(\bar b b\bar c c)$ and soft 
coupling of $(\bar b c)$ bound state \cite{hadr}. In the first stage of 
description, the hard subprocess can be reliably calculated in the framework 
of QCD perturbation theory, while in the second stage the quark binding in 
the heavy quarkonium can be described in the nonrelativistic potential model 
assigned to the $(\bar b c)$-pair rest system. The latter means that
one performs the integration of the final quark state over 
the quarkonium wave function in the momentum space. Since the relative 
quark velocity inside the meson is close to zero, the perturbative matrix 
element can be expanded in series over the relative quark momentum, 
which is low in comparison with the quark masses determining the scale 
of virtualities and energies in the matrix element. 
In the leading approximation one considers only the first nonzero 
term of such expansion, so that for the $S$-wave states the 
matrix element of the parton subprocess for the $B_c$ production is expressed 
through the perturbative matrix element for the production of four heavy 
quarks $(gg\to\bar b b \bar c c)$ with the corresponding projection to
the vector or pseudoscalar spin state of $(\bar b c)$-system, 
which is the color singlet, and through the factor of radial wave function 
at the origin, $R_{nS}(0)$, for the given quarkonium. 
The perturbative matrix element is calculated for the $(\bar b c)$ state, 
where the quarks move with the same 
velocity, i.e. one neglects the relative motion of $\bar b$ and $c$.

For the $P$-wave states, the potential model gives the factor in the form of 
first derivative of the radial wave function at the origin, $R'_{nL}(0)$.
In the perturbative part, one has to calculate the first derivative of the 
matrix element over the relative quark momentum at the point, where the 
velocities of quarks, entering the quarkonium, equal each to other.

Thus, in addition to the heavy quark masses, the values of $R_{nS}(0)$, 
$R'_{nL}(0)$ and $\alpha_s$ are the parameters of calculation for the
partonic production of $B_c$ meson.
In calculations we use the wave function parameters equal to the values
shown in table X and $R'_{2P}(0)=0.50$ GeV$^{3/2}$.
The value of $R(0)$ can be related with the leptonic 
constant, $\tilde f$, so that we have
$$
\tilde f_{1S} = 0.47\;\mbox{GeV,}\;\;\;
\tilde f_{2S} = 0.32\;\mbox{GeV.}\;\;\; \mbox{and}\;\;\;
\tilde f_{n} = \sqrt{\frac{3}{\pi M_{n}}} R_{nS}(0)\;.
$$

In the approximation of the weak quark binding inside 
the meson one has $M_{B_c}= m_b+m_c$,
so that the performable phase space in calculations is close to physical one
at the choices of
$m_b=4.8$ GeV, $m_c=1.5$ GeV for the $1S$-state,
$m_b=5.1$ GeV, $m_c=1.8$ GeV for the $2S$-state,
$m_b=5.0$ GeV, $m_c=1.7$ GeV for the $2P$-state. 

At large transverse momenta of the $B_c$ meson,  $p_T\gg M_{B_c}$,
the production mechanism enters the regime of $\bar b$-quark fragmentation 
(see fig. 1), so that the scale determining the QCD coupling constant in hard 
$\bar b b$ production is given by $\mu^2_{\bar b b}\sim M^2_{B_c}+p^2_T$,
and in the hard fragmentation production of the additional pair of heavy quarks
$\bar c c$ we get $\mu_{\bar c c} \sim m_c$. This scale choice is
caused by the high order corrections of perturbation theory to the 
hard gluon propagators, where the summing of logarithms over the virtualities
leads to the pointed $\mu$ values. Therefore, the normalization of matrix 
element is determined by the value of
$\alpha_s(\mu_{\bar b b})\alpha_s(\mu_{\bar c c})\approx 0.18 \cdot 0.28$. 
In calculations we use the single combined value of $\alpha_s=0.22$.

\vspace*{3mm}
The parton subprocess of gluon-gluon fusion 
$gg\to B_c^+ +b +\bar c$ dominates in the hadron-hadron production of 
$B_c$ mesons. In the leading approximation of QCD perturbation theory it 
requires  the calculation of 36 diagrams in the fourth order over the 
$\alpha_s$ coupling constant. 

By the general theorem on factorization,  it is clear that
at high transverse momenta the fragmentation of the heavier quark
$Q\to (Q\bar q) + q$ must dominate. It is described by the factorized 
formula
\begin{equation}
\frac{d\sigma}{dp_T} = \int \frac{d\hat \sigma(\mu; gg\to Q\bar Q)}
{dk_T}|_{k_T=p_T/x}\cdot D^{Q\to (Q\bar q)}(x;\mu)\; 
\frac{dx}{x}\;,
\label{one}
\end{equation}
where $\mu$ is the factorization scale, 
$d\hat\sigma/dk_T$ is the cross-section
for the gluon-gluon production of quarks $Q+\bar Q$, $D$ is the fragmentation
function. 

The calculation for the complete set of diagrams of the
$O(\alpha_s^4)$-contribution \cite{hadr}
allows one to determine a value of the transverse momentum $p_T^{min}$, 
which is the low boundary of the region where
the subprocess of gluon-gluon $B_c$-meson production enters 
the regime of factorization for the hard production of $b\bar b$-pair and 
the subsequent fragmentation of $\bar b$-quark into the bound 
$(\bar b c)$-state, as it follows from the theorem on the
factorization of the hard processes in the perturbative QCD. 

\begin{figure}[p]
\hspace*{1cm}$d\hat\sigma/dp_T$, nb/GeV\\
\vspace*{-5mm}
\begin{center}
\hspace*{-10mm}
\epsfxsize=14cm \epsfbox{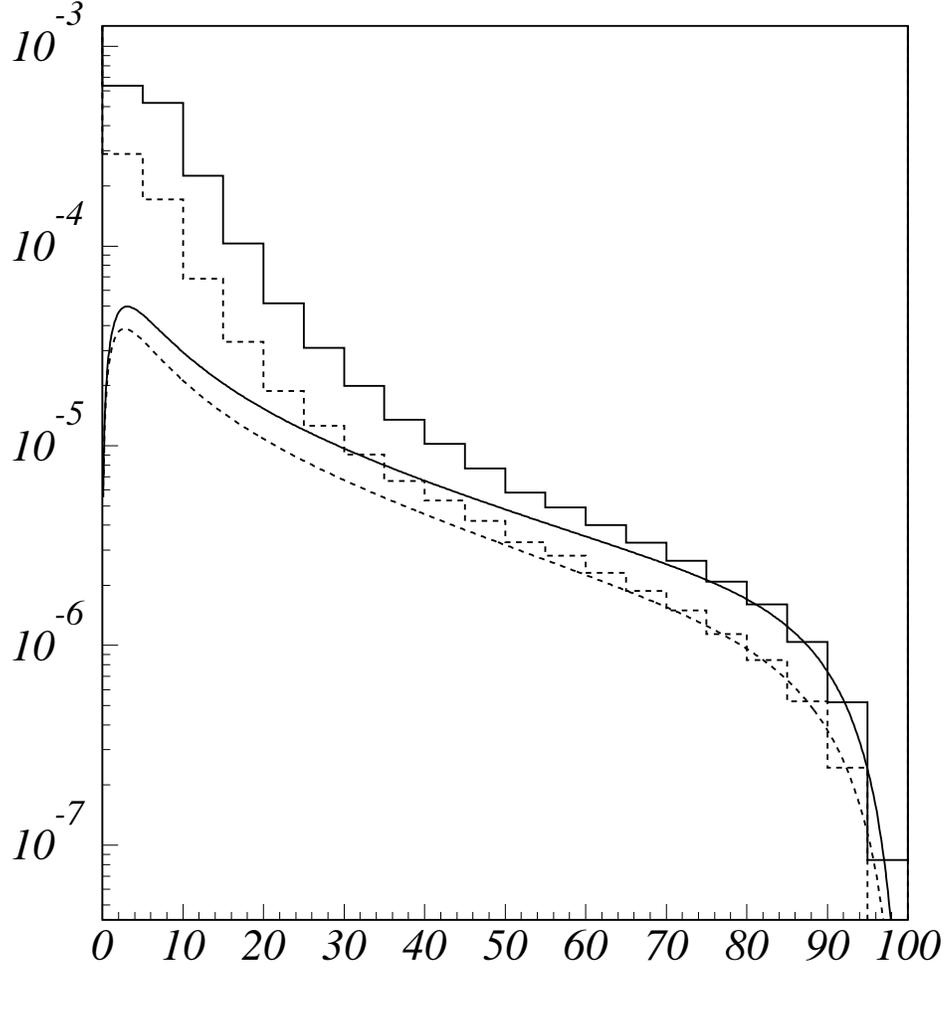}
\end{center}
\vspace*{-13mm}
{\hfill GeV\hspace*{8mm}}
\caption{The differential cross-section for the $B_c^{(*)}$ meson production in
gluon-gluon collisions as calculated in the perturbative QCD over the complete
set of diagrams in the $O(\alpha_s^4)$ order at 200 GeV.
The dashed and solid histograms present the pseudoscalar and vector states, 
respectively, in comparison with the corresponding results of fragmentation model
shown by the smooth curves.}
\label{fig1}
\end{figure}

The $p_T^{min}$ value turns out to be much greater than the $M_{B_c}$ mass, 
so that the dominant contribution into the total cross-section of
gluon-gluon $B_c$-production is given by the diagrams of 
nonfragmentational type, i.e. by the recombination of heavy quarks. 
Furthermore, the convolution of 
the parton cross-section with the gluon distributions inside the initial 
hadrons leads to the suppression of contributions at large transverse 
momenta, as well as the subprocesses with large energy in the system of parton 
mass centre, so that the main contribution into the total cross-section of 
hadronic $B_c$-production is given by the region of energies less or
comparable to the $B_c$-meson mass, where the fragmentation model can not be 
applied by its construction. Therefore, one must perform the calculations 
with the account for all contributions in the given order under consideration
in the region near the threshold.

The large numeric value of 
$p_T^{\rm min}$ means that the majority of events of the 
hadronic $B_c^{(*)}$-production does not certainly allow the description in
the framework of the fragmentation model. This conclusion looks more evident, 
if one considers the $B_c$-meson spectrum over the energy.

The basic part of events for the gluon-gluon production of
$B_c$ is accumulated in the region of low $z$ close to 0, where the 
recombination  dominates.
One can draw the conclusion on the  essential 
destructive interference in the region of $z$ close to 1
and $p_T < p_T^{\mbox{min}}$ , for the pseudoscalar
state.

We have considered in detail the contributions of
each diagram in the region of $z\to 1$. In the covariant Feynman gauge the
diagrams of the gluon-gluon production of $Q+\bar Q$ with the subsequent
$Q\to (Q\bar q)$ fragmentation dominate, as well as the diagrams
when the $q\bar q$ pair is produced in the region
of the initial gluon splitting. 
However, the contribution of the latter diagrams
leads to the destructive interference 
with the fragmentation amplitude, and this results in the "reduction"
of the production cross-section in the region of $z$ close to 1.
In the axial gauge with the vector
$n^\mu=p^\mu_{\bar Q}$ this effect of the interference still manifests itself
brighter, since the diagrams like the splitting of gluons 
dominate by several orders of magnitude over the fragmentation, but the
destructive interference results in the cancellation of such extremely
large contributions. This interference is caused by the 
nonabelian nature of QCD, i.e., by the presence of the gluon self-action 
vertices.

The using of CTEQ5L parameterization for the structure functions
of nucleon \cite{cteq} leads to the total hadronic cross-sections 
for the $B_c$ mesons of about 0.8 $\mu$b that accepts contributions from:
\begin{displaymath}
\begin{array}{cccc}
 1S_0  &  1S_1  & 2S_0 & 2S_1 \\
 0.19\;\mu\mbox{b} & 0.47\;\mu\mbox{b} & 0.05\;\mu\mbox{b} & 0.11\;\mu\mbox{b}
 \end{array}
\end{displaymath}

After the summing over the
different spin states, the total cross-sections for
the production of $P$-wave levels is equal to 7\%  of the $S$-state
cross-section.

At LHC with the luminosity ${\cal L}=10^{34}$ cm$^{-2}$s$^{-1}$ and
$\sqrt{s}=14$ TeV one could expect 
$4.5\cdot 10^{10}\;\; B_c^+$ events per year.

Nevertheless, the P-wave states could be of a particular interest due to their
radiative decays with relatively energetic photons (around 500 MeV in the
$B_c$ rest frame). For P-wave states, the leading color-singlet matrix element
and the leading color-octet matrix elements are both suppressed by a factor of
$v^2$ (relative velocity of the charmed quark) relative to the color-singlet
matrix element for S-wave that can enhance the P-vawe contribution.

\begin{figure}[hbt]
\begin{center}
\epsfxsize=7cm \epsfbox{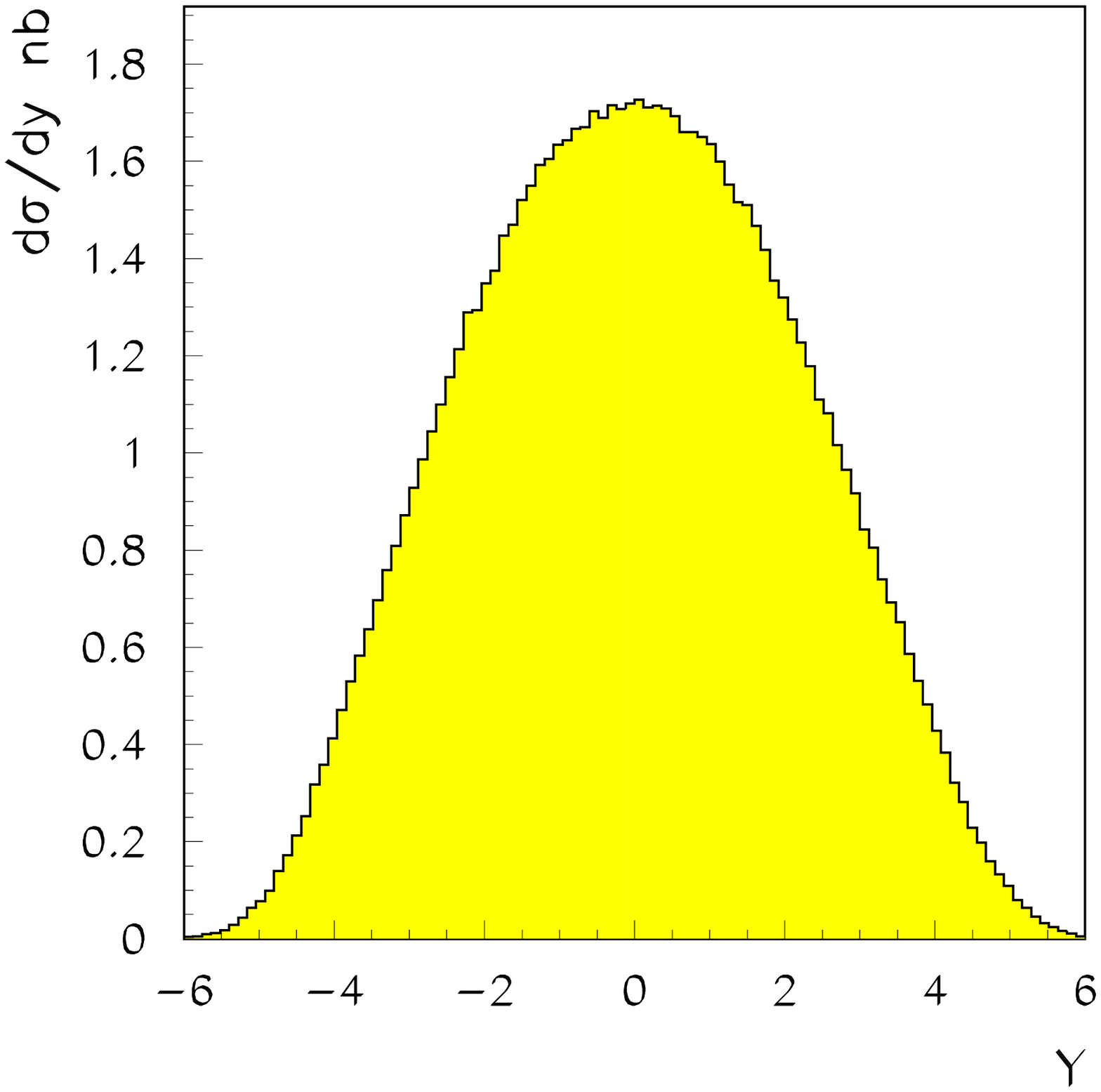}~~
\epsfxsize=7cm \epsfbox{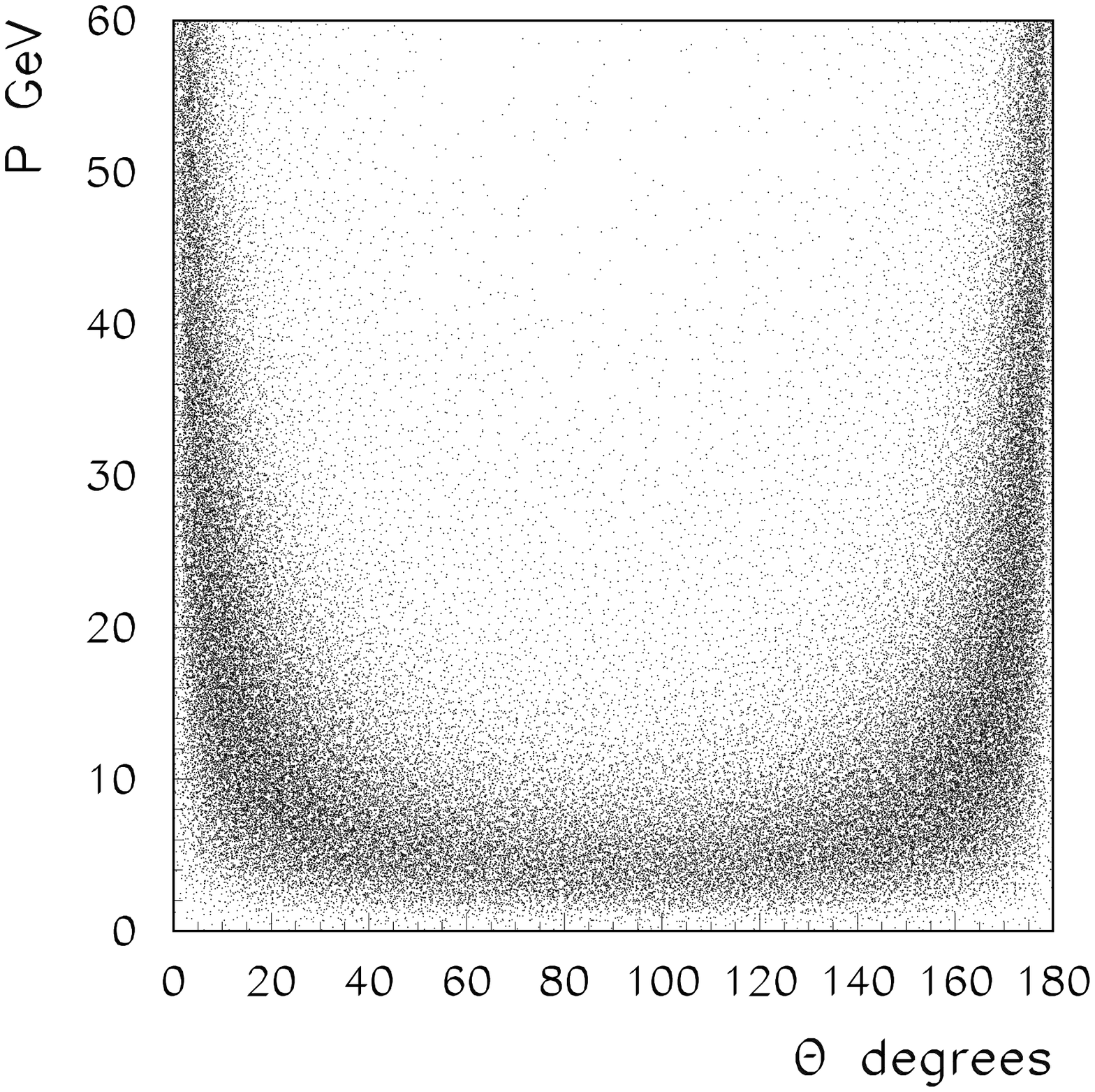}~~
\end{center}
\caption{$d\sigma/dy$ (left) and 
$B_c$ momentum vs. angle (right) for the $B^{+}_c(1S_0)$ at LHC energy.}
\label{lhcbfig}
\end{figure}

In the Figure \ref{lhcbfig} 
the $d\sigma/dy$ distribution is presented.
The $y$ distribution shows the maximum in the central
region however, considereing the experimental observability of the $B_c$ 
states, one
should care about the momentum of the meson to ensure the reasonable
$\gamma$-factor for $B_c$ and visible separation of $B_c$ decay vertex from the
primary one.

The right part of the same Figure 
shows the dependence of the $B_c$ momentum on the angle. One can
see that the central region is dominated by low-momentum mesons, which makes
the observation of these states in the central region quite difficult task. 
On the other hand, the
forward-backward (LHCb acceptance) regions are dominated by very energetic 
mesons.

In the region of the LHCb acceptance ($\Theta < 17^o$) the expected number of
events with the nominal luminosity of $2\times10^{32}$ is about $10^9$ per year.
Taking the value of $\sim 0.1$ as 
an approximation for the reconstruction efficiency of the decay 
$B_c\to J/\Psi \pi \to \mu\mu\pi$, for example, one gets the total amount of 
$2\times 10^4$ reconstructed events per year.

The topology of the events with $B_c$-mesons production is somewhat specific due
to extreme kinematics which, particularly, is responsible for the enhancement of
the forward-backward regions. The main feature of these events topology consists
in the strong corellation in the direction of $B_c$ and associated $D$- and
$b$-mesons momenta. One should expect an associated production of all three 
heavy mesons in the same hemisphere and, moreover, in sufficiently narrow cone.

\section{Conclusions}

The family of $(\bar b c)$ mesons contains 16 narrow states. The S-wave
ones will be produced in $pp$ collisions at LHC energies with relatively large
cross-sections, $\sim 0.1\;\mu\mbox{b}$. The total cross-section of the $B_c$
production, taking into accout the cascade decays of the narrow excited states,
can be as high as $\sim 1\;\mu\mbox{b}$. This value is more than order of
magnitude larger than at
Tevatron energy.

With the total luminosity of about  
{$\cal L$} $=10^{34}\;\mbox{cm}^{-2}\mbox{sec}^{-1}$, one could expect the total
amount of $B_c$ mesons produced of the order of $N_{B_c} \sim 5\cdot 10^{10}$
per year.

The forward-backward regions of the $B_c$-mesons production are more favourable
in the view of experimental observation due to the strong Lorentz boost of the
initial parton system.  One can expect $\sim 10^9$ $B_c$
events per year inside LHCb acceptance. This amount is quite sufficient to study
the spectroscopy and various decay modes, as well as the life-time of the
ground state.

The inclusive decay mode $B_c \to J/\Psi X$ has a branching of about 17\%, in
comparison with $\sim$1\%\ in $B_{d,s}$-mesons decays. These channels produce a
very well visible signatures and one could expect $\sim (2-4)\cdot 10^4$ events 
in the decay mode $B_c\to J/\Psi \pi \to \mu\mu\pi$ and $\sim 8\cdot 10^4$
events in the $B_c \to J/\Psi \mu\nu \to \mu\mu\mu\nu$ mode. The
Cabbibo-suppressed mode $B_c\to J/\Psi K \to \mu\mu K$ can be observed at the
level of $\sim 10^3$ events.

The most probable $b$-spectator decay modes are saturated by the two-body
decays $B_c \to B_s^{(*)} \pi^{\pm}$ and $B_c \to B_s^{(*)} \rho^{\pm}$, that
makes these channels quite interesting for the study of 
the $c$-quark decays in the
$B_c$-meson. The estimated yield of the reconstructed events could be $\sim
10^3$ events per year in the $B_c \to B_s \pi^{\pm}$ mode. Approximately the
same amount of events can be observed
in other modes with $B_s$ in the final state, although with
worse background conditions.

Rare decay modes, e.g. $B_c \to D^{\pm} D^0$, could be interesting in view 
of CP-violation studies, however the preliminary estimations of the detection
efficiencies and branchings involved are not too optimistic and 
additional studies of the reconstruction and selection procedures are required.

\vspace*{0.5cm}

Authors are very grateful fot the contribution of Alexandr Berezhnoj to this
work. We also would like to thank LHCb collaboration for warm hospitality and
IHEP group (Protvino) in LHCb that provides a support for this work.

\end{document}